# Numerical Modeling of Stress Corrosion Cracking in Steel Structures with Phase Field Method

M. Askari[1], P. Broumand*[1], M. Javidi[2]

[1] Department of Civil and Environmental Engineering, Shiraz University, Shiraz, Iran

[2] Department of Material Science Engineering, Shiraz University, Shiraz, Iran

**Abstract**

This study presents a novel coupled mechano-electro-chemical formulation for predicting stress corrosion cracking (SCC) phenomena in steel structures using the phase field method. SCC is a complex damage process that arises from the interaction between mechanical loading and corrosion in a corrosive electrolyte environment. The proposed formulation introduces a new phase-field parameter that aggregates the damage due to mechanical loading and electro-chemical corrosion. To achieve this goal, the internal energies governing the SCC phenomenon are separated into elastic-damage strain energy, the interfacial reaction energy, and energy resulting from changes in corrosion ion concentration. The Allen-Cahn equation is modified to include all energy contributions and calculate the phase field parameter. Furthermore, a specific interfacial kinetic coefficient is introduced to the mechanical energy to take into account corrosion current effects on mechanical properties. The Cahn-Hilliard equation is applied to model the corrosion ion concentration in the domain and the mechanical state of the body is obtained by solving the equilibrium equations. Several numerical examples are presented to validate the robustness and accuracy of the proposed formulation. Finally, the method is applied to predict crack propagation resulting from SCC on two practical engineering problems, yielding promising results.

---

* Corresponding author:
E-mail address: pbroumand@shirazu.ac.ir (P. Broumand)

**1. Introduction**

Stress corrosion cracking (SCC) is commonly defined as crack growth resulting from the combined effects of a corrosive environment and mechanical loading [1, 2]. SCC often leads to brittle fracture and thus requires particular attention [3]. Steel structures are frequently exposed to corrosive environments, such as offshore platforms, pipelines, chemical and petrochemical plants, oil refinery facilities, nuclear power plants, and cooling water systems. Hot spots, where stress concentration is accompanied by moisture entrapment, are particularly susceptible to SCC. This includes weldments and bolted components, which are essential for the overall stability of the structure, such as beam-to-column connections, bracings, base-plates, etc.

From a microscopic perspective, crack growth due to SCC can be either trans-granular or inter-granular and involves various mechanisms, including anodic dissolution, film rupture, and hydrogen embrittlement [3]. Anodic dissolution is the primary electrochemical mechanism of SCC, also known as the active path mechanism [4]. It involves the transformation of metallic atoms into cations and the release of electrons in the anodic reaction. In film rupture, the oxide layer that forms on the crack surface ruptures due to plastic deformation and stress concentration at the crack tip, exposing the bare metal to the electrolyte and resulting in further crack growth. In hydrogen embrittlement, small hydrogen atoms penetrate the host metal crystal lattice, leading to decreased metallic material ductility [5]. Given that SCC is a complex mechano-chemical phenomenon, its modeling in steel structures necessitates sophisticated and efficient analytical and numerical schemes, which are the focus of this research.

There are various numerical simulation approaches available for modeling the stress corrosion cracking (SCC) phenomenon. One approach involves explicitly modeling the crack geometry and applying fracture mechanics principles to simulate crack propagation. Within this category, several researchers have utilized finite element method (FEM)-based cohesive zone models (CZM) to simulate SCC. For example, Ahn et al. [6] investigated hydrogen-assisted cracking in ductile metals using CZM. Their study considered the effects of hydrogen on the void-nucleation and growth mechanisms in the traction-separation law and indicated that the penetration of hydrogen atoms into the metallic crystal lattice accelerates the failure process and reduces material fracture toughness. Scheider et al. [7] modeled hydrogen-assisted SCC using CZM and simulated hydrogen diffusion into the metal lattice by Fick's second law. The authors considered the effect of hydrogen concentration on the material cohesive strength and critical separation. Although this method is simple and efficient and can model complex phenomena like crack branching, it suffers from mesh dependency in the crack growth path and may overestimate the fracture toughness.

Another numerical method, the Extended Finite Element Method (XFEM), has been successfully used to simulate SCC. This method is based on partition of unity concepts and separates interfaces like cracks from the background mesh by enriching the finite element space with specially tailored functions, removing the need for re-meshing during crack propagation [8, 9]. Lee et al. [10] used XFEM to model primary water SCC in nuclear reactors. The proposed method estimated the SCC crack initiation time based on material properties and stress level, and the crack would grow when the SCC crack initiation time was reached. Although this method is computationally efficient and can accurately capture the crack growth path, it requires additional criteria for crack propagation and cannot efficiently handle complex phenomena like crack nucleation, branching, and arrest.

Peridynamics is another numerical method that can handle discontinuities such as cracks, where spatial partial derivatives are undefined and classic continuum mechanics theory fails. This method can be considered a continuum version of molecular dynamics (MD) that takes into account the interaction between material points by defining suitable force functions within an influence radius. De Meo et al. [11] used peridynamics to model SCC, considering the adsorption-induced decohesion mechanism. This method can accurately capture the micro-mechanisms of the fracture and thus does not require extra fracture criteria. It can overcome several limitations of the CZM and XFEM methods; however, it is computationally demanding for practical engineering problems.

In the second approach, the crack geometry is not explicitly modeled and the effects of the discontinuities are implicitly applied to the material properties. The continuum damage mechanics (CDM) is the frontier method in this category. The damage parameter, usually a scalar or tensor field, represents the state of each material point ranging from virgin to fully damaged. The strain equivalence concept is usually used to apply the damage effects on the state of stress and stiffness, and the damage evolution law determines the relation between damage growth rate with stress and strain fields, and history-dependent parameters [12]. Local damage models suffer from strain localization due to material softening, hence, different regularization methods have been proposed to remedy, among which non-local damage models are among the most favorite[13]. Bastos et al. [12] proposed a thermodynamically consistent CDM-based model for SCC in austenitic stainless steel. The authors employed continuum damage plasticity formulation and related the damage evolution to the corrosive environment parameters. Jasra et al. [13] used a modified Lemaitre damage model to numerically simulate SCC due to the effects of temperature and Chloride concentration in austenitic stainless steel.

The study showed that the effect of temperature is more significant than chloride concentration in total number of failed elements, number of cracks initiated and rate of elements failure.

The phase field method (PF) is another technique that has been widely used to model cracks implicitly. Originally proposed for modeling multi-phase problems in material science, Frankfort et al. [14] and Bourdin et al. [15] extended the method to model brittle fracture problems in the early 2000s. Similar to CDM, the PF method determines the state of material (intact or cracked) with a scalar called the phase field parameter which regularizes cracks over a predefined characteristics length. The regularization process suppresses the mesh dependency issues induced by material softening to a great extent. The method is based on the variational form of the Griffith energy criterion in the fracture mechanics which results in a Helmholtz-type governing equation for the phase field parameter; in this equation, the free strain energy is the main contributor to the elastic crack growth [16,17]. Several authors have extended PF to model SCC problems. Stahle et al. [18] used PF to simulate surface roughening due to pitting, conversion of pits into cracks, and crack propagation under simultaneous effects of an aggressive environment and mechanical loading. Lin et al. [19] proposed a coupled mechano-chemical PF scheme to model SCC. The authors studied the pit-to-crack transition and the synergic effect of the corrosive environment, mechanical loading, and initial geometry of pits. Mai et al. [20] presented a PF formulation to model activation and diffusion-controlled pitting and the location of the corrosion interface was implicitly traced by solving the phase-field equation. A chemical free energy function was used to consider the electrolyte-metal interaction, from which, the evolution laws of field variables were derived. Nguyen et al. [21] combined PF with image processing techniques to detect crack initiation in Nickel alloys due to SCC. They also employed Fick's second law to model material dissolution and film rupture at the crack-tip zone and considered their effects on the material fracture toughness and thus, crack propagation. Recently, Cui et al. [22] proposed a coupled mechano-electro-chemical PF technique for dissolution-driven cracking; the method can capture pitting corrosion, SCC, and pit-to-crack transition. The film rupture-dissolution-passivation mechanism was modeled by considering the critical plastic strain in the crack tip area and a Tafel like equation for corrosion current density. The dissolution of metal into the electrolyte and corresponding moving interface was modeled by the phase-field and metal concentration parameters. The electrochemical free energy is considered the main driving force of the corrosion which is affected by mechanical loading and it is decomposed into the bulk free energy of the material and interfacial free energy. The variational form of the total energy leads to three sets of governing equations: static equilibrium for solid body, Allen-Cahn [23] equation for the phase-field parameter, and Cahn-Hilliard [24] equation for metal concentration. Although the model appealingly simulates SCC and can handle complex mechanisms, it neglects the mechanical damage due to mechanical free strain energy.

In this research, a coupled mechano-electro-chemical formulation based on the PF method is proposed to model SCC problems. Unlike existing formulations in the literature, the proposed formulation can operate in inert environments under pure mechanical effects, in aggressive environments under pure chemical effects, and in aggressive environments under simultaneous mechanical and chemical effects. In this respect, the phase field parameter is defined as the summation of the mechanical and chemical damages; the total free energy of the system is enhanced to comprise both the mechanical bulk material and crack surface energies, and the electro-chemical bulk material and the interfacial free energies. The Allen-Cahn equation is modified accordingly to consider the above changes in the phase-field governing equation. The Cahn-Hilliard equation is used to calculate the concentration of the metal, and the effects of the phase field damage parameter on the mechanical behavior is applied in the equilibrium equations. The structure of the paper is as follows: the theoretical background on the SCC phenomenon is presented in section 2; the proposed model and the implementational aspects in the framework of COMSOL-Multiphysics software are explained thoroughly in section 3. Finally, section 4 is devoted to model verification and parametric studies on two- and three-dimensional numerical problems.

**2. Theoretical background**

**2.1. Corrosion damage**

All electrochemical galvanic cells consist of two electrodes, a cathode and an anode, and an electrolyte solution. The electron transfer process is carried out in the electrodes, and the ion transfer occurs in the electrolyte. In corrosion, the base metal can play the role of anode and cathode simultaneously. Redox reactions, i.e. oxidation (loss of electrons) and reduction (gain of electrons), occur in the anode and cathode, respectively [25].

The galvanic cell forms an electric circuit and Tafel's law can be used to express the current density of the electrochemical reaction based on the overpotential of the system. This equation can be applied to irreversible reactions in corrosion as below [26]:

$$i_a = i_0 \exp\frac{\eta}{b} \tag{1}$$

where $b$ is the slope of the Tafel curve, $\eta$ is the overpotential applied to the system, $i_0$ is the exchange current density, and $i_a$ is the corrosion current density. In classical electrochemistry, the velocity of the corrosion front, i.e. the rate at which anodic dissolution occurs, is determined based on the concentration of metal ions in the corrosion interface. If the concentration of the ions is lower than the saturation concentration, the process is known as activation-controlled. When the concentration of ions reaches the saturation concentration, the rate of ion transport at the boundary is controlled by the saturation concentration, and the process is called diffusion-controlled [27]. The velocity of the corrosion front in activation-controlled and diffusion-controlled states are modeled by Faraday's second law and Rankine–Hugoniot relationship, respectively. Following the work of Cui et al. [22], the moving interface boundary conditions can be enforced numerically by employing the phase field method and the sharp corrosion interface can be smeared over a characteristic length $l$. For this purpose, a scalar phase-field parameter, $\phi$, is defined which represents the intact metal when $\varphi = 0$, and the corroded metal (i.e. electrolyte solution) when $\varphi = 1$. To express the release of metal ions due to anodic dissolution, normalized concentration, $c$ is defined as [28]:

$$c = \frac{c^{ion}}{c_{solid}^{alloy}} \tag{2}$$

Where $c^{ion}$ and $c_{solid}^{alloy}$ are the overall ionic concentration in the interface and metal atoms in the solid phases respectively. The normalized concentration of metal atoms ($c_S$) and the normalized concentration of ions in the electrolyte ($c_L$) are expressed as follows [28]:

$$c_S = \frac{c_S^{alloy}}{c_{solid}^{alloy}}, \quad c_L = \frac{c_l^{solution}}{c_{solid}^{alloy}} \tag{3}$$

Where $c_S^{alloy}$ and $c_l^{solution}$ are the real concentration values of metal and electrolyte respectively. Different functions are used to express the free energy density function in the electrochemical process. The Kim-Kim-Suzuki (KKS) model [29] is used in this research to represent the free energy density in the phase field method; in this model, the concentration of each point of the system is a combination of two phases with different concentrations and the same chemical potential as follows:

$$c = h_c(\varphi)c_S + [1 - h_c(\varphi)]c_L \tag{4}$$

$$\frac{\partial f_s(c_S)}{\partial c_S} = \frac{\partial f_l(c_L)}{\partial c_L} \tag{5}$$

Where $c_{se}$ and $c_{Le}$ express the normalized equilibrium concentrations for solid and electrolyte phases, $f_s(c_S)$ and $f_l(c_L)$ are the free energy density functions of solid and electrolyte phases, and $h_c(\varphi)$ is a continuous interpolator function based on the phase field variable which expresses the state of the system at each time of the reaction such that $h_c(0) = 1$, $h_c(1) = 0$, $h'_c(0) = 0$, $h'_c(1) = 0$, $f_s(c_s) = A(c_s - c_{se})^2$, and $f_L(c_L) = A(c_L - c_{Le})^2$. Thus, the chemical free energy density function is expressed as follows [20]:

$$f_0(c, \varphi) = [h_c(\varphi)]f_s(c_s) + [1 - h_c(\varphi)]f_l(c_L) + wg(\varphi) \tag{6}$$

In this equation, $g(\varphi) = \varphi^2(1-\varphi)^2$ is the double-well potential, $w = \frac{4\sqrt{2}\,\Upsilon \alpha^*}{\ell}$ is the height of the potential function, $\Upsilon$ is the boundary energy, $\alpha^* = 2.94$, $\ell$ is the thickness of interface and $A$ is the curvature of the energy density function [22]. Combining the above equations yields the KKS form of the chemical free energy density function as below:

$$f_0(c, \varphi) = A[c - h_c(\varphi)(c_{se} - c_{Le}) - c_{Le}]^2 + wg(\varphi) \tag{7}$$

The total chemical energy ($F^\mathbb{C}$) can be expressed as the sum of the bulk energy ($F^c$) and the interface energy ($F^\varphi$) as follows [30]:

$$F^\mathbb{C}[c, \varphi] = F^c + F^\varphi = \int_\Omega [f_0(c.\varphi) + \tfrac{1}{2}\alpha_c(\nabla c)^2 + \tfrac{1}{2}\alpha_\varphi(\nabla \varphi)^2]dv \tag{8}$$

where $\alpha_c$ and $\alpha_\varphi$ are the energy coefficients for the concentration at the interfaces. Considering the problem of mass diffusion in the corrosion process, the Cahn-Hilliard and Allen-Cahn equations can be used to express the rate of concentration change and phase-field damage, respectively [20]:

$$\frac{\partial c(x,t)}{\partial t} = -\nabla \cdot \boldsymbol{j} = \nabla \cdot M\left(\nabla \frac{\partial F}{\partial c}\right) \quad \text{where } M = \frac{D}{2A} \tag{9}$$

$$\frac{\partial \varphi(x,t)}{\partial t} = -\mathbb{L}\frac{\delta F}{\delta \varphi} \tag{10}$$

where $\mathbb{L}$ is the kinetic coefficient of the interface which expresses the rate of change of the boundary from solid to electrolyte phase, and $D$ represents the diffusion coefficient. The Allen-Cahn equation is not conservative, i.e. the whole domain material can be transformed from virgin solid to corroded substance, and it has second-order spatial derivatives which makes it suitable to express the interfacial energy in numerical methods; this is in contrast to the Cahn-Hilliard equation which is conservative and has fourth-order spatial derivative [31]. Thus, in this work, similar to reference [20], the Allen-Cahn equation is used to simulate the energy changes at the boundaries, i.e. $\alpha_c = 0$ and $\alpha_\varphi$ is equal to $\frac{2\sqrt{2}\,\Upsilon \ell}{\alpha^*}$.

**2.2. Mechanical damage**

Based on the Griffith's energy criterion in the fracture mechanics, the total mechanical internal energy of a cracked body can be expressed as the sum of the elastic energy stored in the bulk material ($E_u$) and the energy dissipated due to the crack formation ($E_s$) [32],

$$F^u(\mathbf{u}, \Gamma) = E_u(\mathbf{u}, \Gamma) + E_s(\mathbf{u}, \Gamma) = \int_\Omega f_u[\boldsymbol{\varepsilon}(\mathbf{u}), \varphi] dv + \int_\Omega G_c \gamma(\varphi, \nabla\varphi) dv \qquad (11)$$

where **u** is the displacement field, $\Gamma$ is the crack surface, $f_u[\boldsymbol{\varepsilon}(\mathbf{u}), \varphi]$ is the elastic energy density function, $G_c$ is the fracture toughness, and $\gamma(\varphi, \nabla\varphi)$ is the crack density per volume of the solid body [17,21]. The mechanical phase field variable $\varphi$ simulates the state of the crack damage from a macroscopic point of view. Using the concepts of the calculus of variations, the crack density function can be expressed in terms of the phase field parameter as,

$$\gamma(\varphi, \nabla\varphi) = \frac{1}{2l} \varphi^2 + \frac{l}{2} (\nabla\varphi)^2 \qquad (12)$$

The elastic strain energy density function is decomposed to tensile, $\psi^+(\boldsymbol{\varepsilon})$, and compressive parts, $\psi^-(\boldsymbol{\varepsilon})$, respectively, to distinguish the different crack behaviors in tension and compression and crack closure effects in compression, as below

$$f_u = [h_m(\varphi) + \kappa] \psi^+[\boldsymbol{\varepsilon}(\mathbf{u})] + \psi^-[\boldsymbol{\varepsilon}(\mathbf{u})] \qquad (13)$$

$$\psi^+(\boldsymbol{\varepsilon}) = \frac{\lambda}{2}[Tr(\boldsymbol{\varepsilon})_+]^2 + \mu\, Tr[(\boldsymbol{\varepsilon})_+^2] \qquad (14)$$

$$\psi^-(\boldsymbol{\varepsilon}) = \frac{\lambda}{2}[Tr(\boldsymbol{\varepsilon})_-]^2 + \mu\, Tr[(\boldsymbol{\varepsilon})_-^{\,2}] \tag{15}$$

where $(\boldsymbol{\varepsilon})_+$ and $(\boldsymbol{\varepsilon})_-$ represent the tensile and compressive parts of the strain tensor, and $\lambda$ and $\mu$ are the Lamé's elasticity constants. $h_m(\varphi)$ is the phase field degradation function determined such that $h_m(0) = 1, h_m(1) = 0$ and $h'_m(1) = 0$, and $\kappa$ is a small constant which is meant to prevent the convergence issues and numerical instabilities in the case of the fully damaged zone. The phase field parameter also affects the constitutive equation as follows:

$$\boldsymbol{\sigma} = h_m(\varphi)\{\lambda[Tr(\boldsymbol{\varepsilon})_+] + 2\mu\, Tr[(\boldsymbol{\varepsilon})_+]\} + \{\lambda[Tr(\boldsymbol{\varepsilon})_-] + 2\mu\, Tr[(\boldsymbol{\varepsilon})_-]\} \tag{16}$$

To ensure that the crack healing in compression is prevented, the strain energy density function history parameter $\mathbb{H}$ is defined as below:

$$\mathbb{H}(\boldsymbol{x}, t) = \max_{0 \leq t_0 \leq t}\left\{\frac{l}{G_c}\psi^+(\boldsymbol{x}, t_0)\right\} \tag{17}$$

The minimization of the total mechanical energy yields the equilibrium equations and the Helmholtz equation for the mechanical phase-field damage as follows (for more details see [17])

$$\nabla \cdot \boldsymbol{\sigma} + \rho\boldsymbol{b} = \boldsymbol{0} \tag{18}$$

$$\phi - l^2 \Delta\phi = -h'_m \mathbb{H} \tag{19}$$

## 3. Phase Field Modelling of SCC

This study assumes that the phase field parameter, $\varphi$, aggregates mechanical and electrochemical damages (cracks). This assumption is based on the following reasoning: when a crack grows mechanically, the new fracture surfaces are exposed to an electrolyte solution, resulting in corrosion as if the material had been previously corroded. Although fluid movement in the fractures takes time, the slow rate of subsequent corrosion allows for the assumption that it occur simultaneously with crack growth. Similarly, when a crack grows due to corrosion, the material fails, and its mechanical properties degrade as if the crack had grown mechanically. In both cases, the material can no longer bear applied loads, leading to identical macroscale continuum degradation behavior. Therefore, the total energy of the system is divided into mechanical, chemical, and interfacial energies.

$$F = F^u + F^{\mathbb{C}} = E_u + E_s + F^c + F^{\varphi} \tag{20}$$

By inclusion of the energy definition into the Allen-Cahn equation, the phase-field damage due to SCC can also be obtained,

$$\frac{\partial \varphi}{\partial t}(x,t) = -\mathbb{L}\frac{\delta F(u,\varphi,c)}{\delta \varphi} = -\mathbb{L}\left(\frac{\partial F^u}{\partial \varphi} + \frac{\partial F^{\varphi}}{\partial \varphi} + \frac{\partial F^c}{\partial \varphi} - \nabla \cdot \left(\frac{\partial F^u}{\partial \nabla \varphi} + \frac{\partial F^{\varphi}}{\partial \nabla \varphi} + \frac{\partial F^c}{\partial \nabla \varphi}\right)\right) \tag{21}$$

Expanding the equation yields,

$$\frac{\partial \varphi}{\partial t}(x,t) = -\mathbb{L}\{-h'_m(\varphi)\mathbb{H} - \varphi - 2A[c - h(\varphi)(c_{se} - c_{Le}) - c_{Le}](c_{se} - c_{Le})h'_c(\varphi) + w.g'(\varphi) + l^2\Delta\varphi - \alpha_{\varphi}\Delta\varphi\} \tag{22}$$

The above formulation can be enhanced by separating the mechanical and electrochemical energy contributions to provide more degrees of freedom in the prediction of the SCC phenomenon. To this goal, different interface kinematic coefficients are applied to the mechanical and chemical parts in the right-hand side of the equation as below

$$\frac{\partial \varphi}{\partial t}(x,t) = -L_{cm}[-h'_m(\varphi)\mathbb{H} - \varphi + l^2\Delta\varphi] - L_{scc}\{-2A[c - h_c(\varphi)(c_{se} - c_{Le}) - c_{se}](c_{se} - c_{Le})h'_c(\varphi) + w.g'(\varphi) - \alpha_{\varphi}\Delta\varphi\} \tag{23}$$

where $L_{cm}$ is the mechanical interface kinetic coefficient and $L_{scc}$ is the SCC interface kinetic coefficient. These coefficients are explained in detail in the next sections. This formulation is now capable to model the separate effects of mechanical or chemical damages by taking $L_{SCC} =$ 0 or $L_{cm} = 0$, respectively; the simultaneous interactive effects of both mechanisms can also be considered by choosing appropriate values for these coefficients.

The modified Cahn-Hilliard equation can also be expressed by substituting the above energy definition in equation (9) as

$$\frac{\partial c}{\partial t}(x, t) = D\Delta[c - h_c(\varphi)(c_{se} - c_{Le}) - c_{Le}] \tag{24}$$

### 3.1. Interface Kinetic Coefficient (*L*)

In the phase field method, the phase change rate in the electrochemical state is expressed by the Allen-Cahn Equation, and the velocity of the boundary ($\frac{\partial \varphi}{\partial t}$) is controlled by the interface kinetic coefficient parameter (*L*). In the state of activation control corrosion, this velocity is determined based on the Faraday's second low, where the corrosion current density has a direct relationship with the corrosion velocity. Mai et al [20] showed that parameter L has a direct relationship with the corrosion current. Therefore, the rate of boundary changes from activation-controlled to diffusion-controlled corrosion is expressed as the change in the interface kinetic coefficient parameter (*L*) relative to overpotential. By using the Tafel's equation, the relationship between the interface kinetic coefficient parameter and the overpotential can be rewritten as below:

$$\eta = b \log \frac{i_a}{i_0} = b \log \frac{L}{L_0} \tag{25}$$

where $L_0$ represents the boundary kinetic coefficient at zero overpotential ($\eta = 0$). By rearranging equation (25), the value of the interface kinetic coefficient parameter is determined for corrosive environments with different corrosion current densities ($i_a$), for the electrochemical effects only, thus [20]:

$$L = i_a \frac{L_0}{i_0} \tag{26}$$

## 3.2. Mechanical Interface Kinetic Coefficient ($L_{cm}$)

The degradation of the mechanical properties of a metal specimen can occur as a result of its reaction with an electrolyte in a corrosive environment. The extent of this degradation is directly proportional to the corrosiveness of the environment, with an increase in electrolyte concentration resulting in a higher corrosion current ($i_a$). The electrochemical interface kinetic parameter exhibits a similar relationship with the corrosion current, whereby an increase in the latter leads to a corresponding increase in the former. In a study by Sadeghi et al. [1] that examined several sample failures due to SCC in a corrosive environment with varying concentrations and constant temperature, it was observed that the failure time of the samples and the yield strength of the material decreased with an increase in electrolyte concentration. The parameter $L$ represents changes in mechanical properties per unit time with respect to changes in the corrosion current density, and multiplying it by the initial properties of the material yields changes in these properties in the elastic state per unit time. To achieve this objective in the current research, the electrochemical interface kinetic coefficient is multiplied by the yield strength ($\sigma_y$), resulting in the parameter $L_{cm}$, as

$$L_{cm} = L . \sigma_y \tag{27}$$

As the corrosion current rate increases, the proposed parameter ($L_{cm}$) also increases, ultimately leading to a decrease in the failure time. It should be noted that there is a similar coefficient in the rate-dependent mechanical phase-field formulation suggested by Mieh et al. [33], known as the kinetic parameter $\eta$.

## 3.3. Interface Kinetic Coefficient ($L_{SCC}$) in SCC

In SCC, the electrochemical reaction rate changes due to the mechanical effects caused by external loading; the $L_{SCC}$ parameter is intended to simulate this effect. To determine this coefficient, the quasi-static stress-strain diagram of a sample specimen in the corresponding corrosive environment with known corrosion current rate is obtained in the laboratory at a constant temperature (see example 4.2). By numerical simulation of the experimental model, the $L_{SCC}$ parameter is then calibrated for the specified corrosion environment with known $L_{cm}$ such that the engineering stress-strain curve coincides well with the experimental results .

### 3.4. The Finite element implementation

In this research, the nonlinear finite element method is used to solve the governing equations of the proposed SCC formulation. To this end, the weak form of the equations is obtained by the methods of the calculus of variations, and corresponding discretized FEM systems of equations are obtained by the Babnov-Galerkin method. The commercial COMSOL-Multiphysics software is then used to implement the proposed formulation. Three different "physics" modules of COMSOL are employed to simulate the governing equations; the "Solid Mechanics" module is used for the mechanical equilibrium equation (18) which specifies the solid materials' mechanical properties, contact constraints, loading, and boundary conditions. The module is modified to include the effects of the degradation function on the mechanical properties of the materials. The electrochemical governing equations, i.e. the Allen-Cahn and Cahn-Hilliard equations as expressed in (23) and (24), are modeled by the "Mathematics" module, through the submodule "Weak form" which solves the weak form of Partial Differential Equations (PDEs). Finally, to solve for the history parameter ($\mathbb{H}$), as stated in Equation (17), the submodule "Ordinary Differential Equation (ODE)" from the "Mathematics" module was used.

## 4. Results and Discussions

In this section, several numerical examples are presented to assess the robustness and efficiency of the proposed formulation for SCC.

### 4.1. SCC in a rectangular steel plate

To verify the implemented formulation in the framework of COMSOL multi-physics, the model presented by Cui et al. [22] is simulated numerically. The problem consists of a rectangular steel plate in contact with a corrosive environment from the upper semicircular hole and subjected to prescribed horizontal displacements from the two side edges. Figure 1, shows the geometry and boundary conditions of the problem. Chemical Dirichlet boundary conditions are $\varphi = 1$, $c = 0$ at the hole boundary, and the initial conditions in the domain are $\varphi = 0$, $c = 1$. The plate is consequently subjected to prescribed displacements of $u^\infty = 0.02, 0.1$ and $0.125\ \mu m$; when the displacement reaches the desired value, it will be kept constant in the rest of the analysis. The mechanical properties of the plate are $E = 190\ GPa$, $v = 0.3$, $\sigma_y = 520\ MPa$, respectively. Since plasticity was also included in reference [22], the von Mises criterion was implemented in the model and the isotropic strain-hardening was considered by the following power law function for the yield stress ($\sigma_y$):

$$\sigma_y = \sigma_{y0}(1 + \frac{E\varepsilon^p}{\sigma_{y0}})^N \tag{28}$$

where $\varepsilon^p$ is the effective plastic strain, $\sigma_{y0}$ is the initial yield stress, and $N$ is the hardening power determined experimentally as 0.067 for this material. The mechanical damage is neglected in this example, i.e. $L_{cm} = 0$, and the value of $L_{SCC}$ is determined by below equation [22]:

$$L_{SCC} = \left[\left(\frac{\varepsilon^p}{\varepsilon^y} + 1\right) \exp\left(\frac{\sigma_h V_m}{RT}\right)\right] L \tag{29}$$

where $\sigma_h$ is the hydrostatic pressure, $\varepsilon^y$ is the effective yield strain, $V_m$ is the molar volume, $R$ is the ideal gas constant, $T$ is the absolute temperature and $L = 0.001 \frac{mm^2}{N.s}$. The summary of electrochemical properties is presented in table 1. The solution time was 900 s.

**Table 1:** Electrochemical properties of the corrosive environment [22].

| Parameter | Value | Unit |
| --- | --- | --- |
| Interface energy ϒ | 53.5 | N/mm |
| characteristic length $l$ | 0.005 | mm |
| Diffusion coefficient $D$ | 8.5 × 10⁻⁴ | mm²/s |
| Interface kinetics coefficient $L$ | 2 × 10⁶ | mm²/(N.s) |
| Free energy density curvature $A$ | 53.5 | N/mm² |
| The average concentration of metal $c_{solid}$ | 143 | mol/L |
| Average saturation concentration $c_{sat}$ | 5.1 | mol/L |

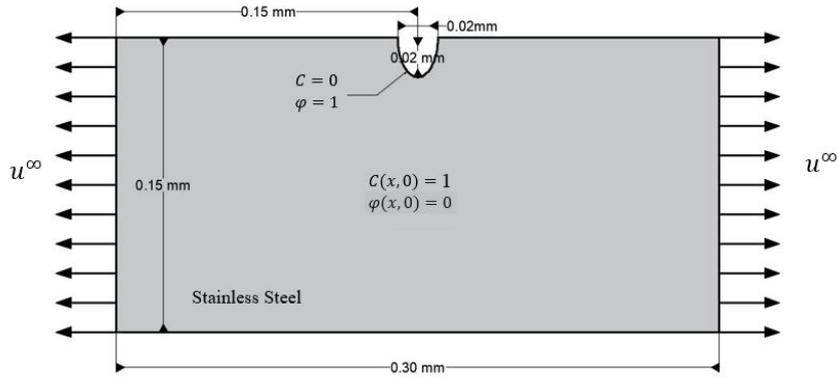

**Fig. 1**. Geometry and boundary conditions of the SCC in rectangular plate problem [22].

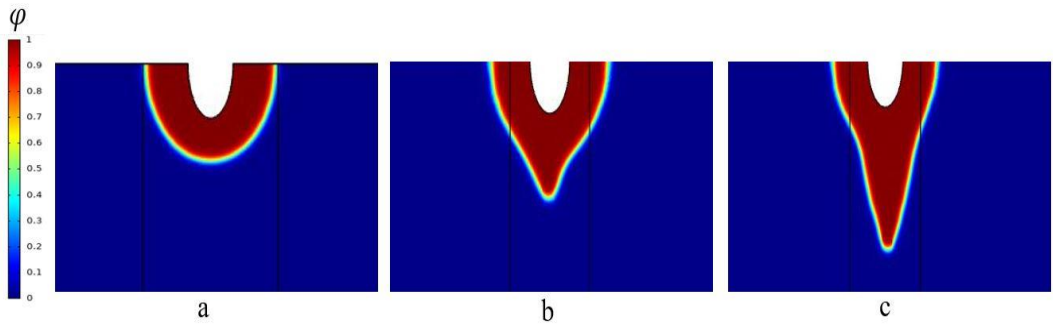

**Fig. 2**. Phase-field parameter contour under different prescribed displacements, (a) $u^\infty = 0.02\ \mu m$, (b) $u^\infty = 0.1\ \mu m$, and (c) $u^\infty = 0.125\ \mu m$.

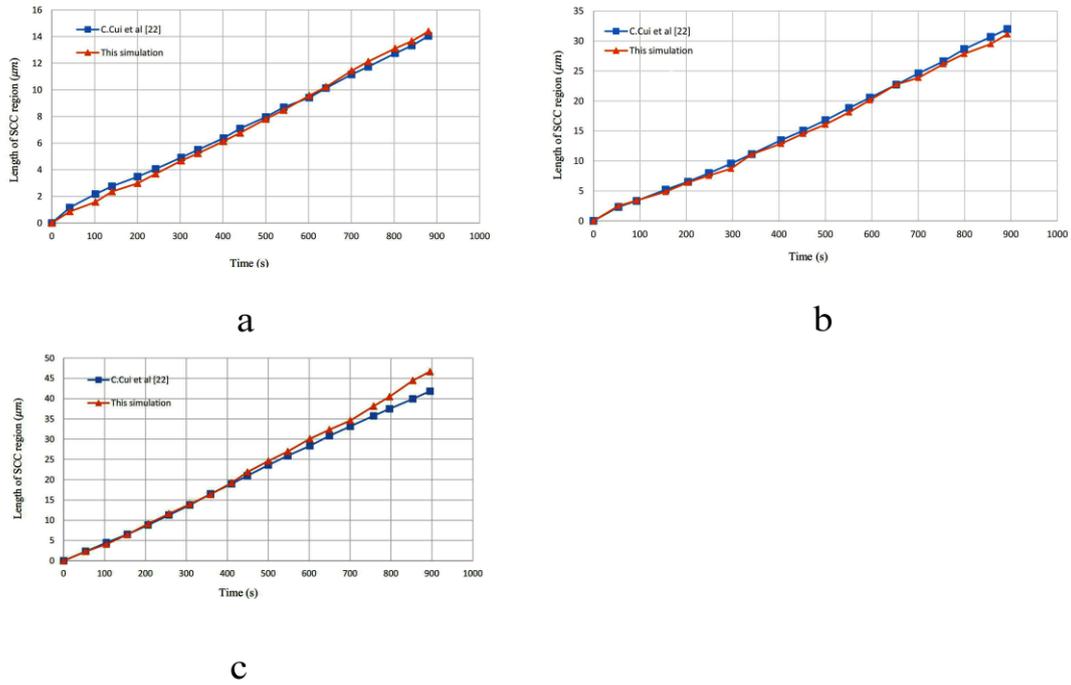

**Fig. 3**. Crack length time histories under different prescribed displacements, (a) $u^\infty = 0.02\ \mu m$, (b) $u^\infty = 0.1\ \mu m$, and (c) $u^\infty = 0.125\ \mu m$.

In the first part, the effects of the mechanical load on the SCC damage are investigated. Figure 2, illustrates the phase filed contours under different prescribed displacements; the results are in good agreement with that of Cui et al [20]. Under small displacements, i.e. $u^\infty = 0\cdot 02\ \mu m$, the electrochemical effects dominate and the damage diffuses uniformly from the hole into the domain. With the increase of the prescribed displacements, the damage obtains a more crack-like appearance and the pit-to-crack phenomenon is observed. In this case, the mechanical effects induce stress concentration at the tip of the hole which accelerates the crack growth in this region.

Figure 3 compares the crack length time histories of Cui et al [20] with this simulation for the three prescribed displacements and the agreement is evident.

### 4.2. SCC in a tensile test

In this example, the calibration process of $L_{scc}$ is explained and the performance of the proposed formulation in the prediction of the SCC failure in an experiment by Sadeghi et al.

[1] is presented. In the experiment, a cylindrical rod sample, as presented in Figure 4, is placed in a cell with corrosive aqueous 1wt.% sodium chloride solution saturated with $CO_2$ at 70 °C. The sample dimensions are $L_1 = 25.4\ mm, D_1 = 3.81\ mm, D_2 = 6.35\ mm$ and $R = 35\ mm$, and it is prepared based on the NACE-TM0198 standard. The rod is made of API 5L X65 steel

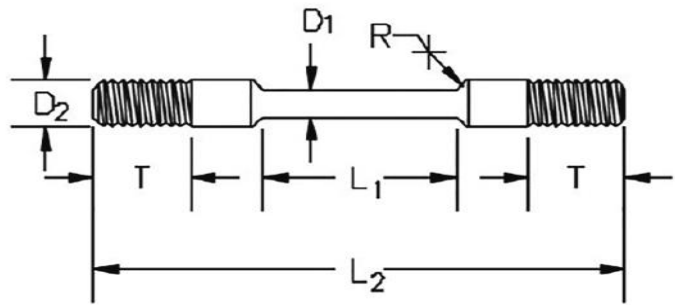

**Fig. 4.** The cylindrical rod sample used for the SCC test [1].

with $E = 200\ GPa$, $\sigma_y = 554\ MPa$, $G_c = 2.7 \times 10^{-3}\ \frac{kN}{mm}$, and $l = 0.0375\ mm$. A tensile test is performed on the sample up to the failure with a constant strain rate of $10^{-5}\ \frac{1}{s}$, and the corresponding stress-strain diagram is obtained. The measured interface kinetic coefficient of the test is $L = 0.0000235\ \frac{mm^2}{N.s}$.

In the numerical simulation, the kinetic interface coefficient is calibrated as $L_{SCC} = 0.000235\ \frac{mm^2}{N.s}$ for which the stress-strain curve has the best agreement with the experimental; using equation (27), $L_{cm} = 0.0130\ \frac{1}{s}$, and other model parameters are kept similar to the previous problem. A perfect-plastic von Mises constitutive model is also used to simulate the plastic behavior of the steel material. To obtain the engineering stress-strain curve, the applied load at each time step, calculated by the integration of the traction forces on the sample's cross-section in COMSOL, is divided by the sample's initial cross-sectional area. The engineering strains are obtained by dividing the gage displacement by the initial gage length, as performed in the experiment.

In the experiment, since the sample is placed in a corrosive environment, the surface protective film dissolves, localized corrosion occurs and small random pits are formed on the surface. These pits are suitable regions for crack incubation under increasing displacement. As cracks propagate from the pits, damage grows and the sample finally fails; due to the highly corrosive condition of the solution, the failure occurs in a quasi-brittle manner [1]. To simulate this condition in the numerical model, an initial random pit, in the form of a cone, with a radius of 0.01 mm and a depth of 0.05 mm is placed on the surface of the sample. This defect is used to model the process of the pit-to-crack conversion. To reduce the computational cost while retaining the accuracy, the SCC formulation is only applied to the middle zone of the sample and a fine mesh is used in this region; the rest of the specimen is modeled with elastic behavior and a coarse mesh.

Figure 5 compares the stress-strain diagrams of the experiment with the numerical simulation; the curve due to the experiment in the inert air environment is also presented to illustrate the pure mechanical behavior of the specimen. As observed, there is an acceptable match between the results in the elastic and plastic regions of the curves. The experimental sample and the numerical model yield at 452 MPa and 462 MPa, and start damaging at 496.8 MPa and 488.8 MPa, respectively. The discrepancy in the failure point can be attributed to the absence of the plastic strain energy in the phase field formulation presented in this research. As shown in Figure 5, when the specimen is exposed to the corrosive environment, the mechanical properties deteriorate, leading to a lower yield strength compared to the intact specimen. The effect of environmental factors on the mechanical properties is captured by the parameter $L_{cm}$ in the formulation, where an increase in the corrosion rate leads to a higher rate of decrease in the mechanical properties

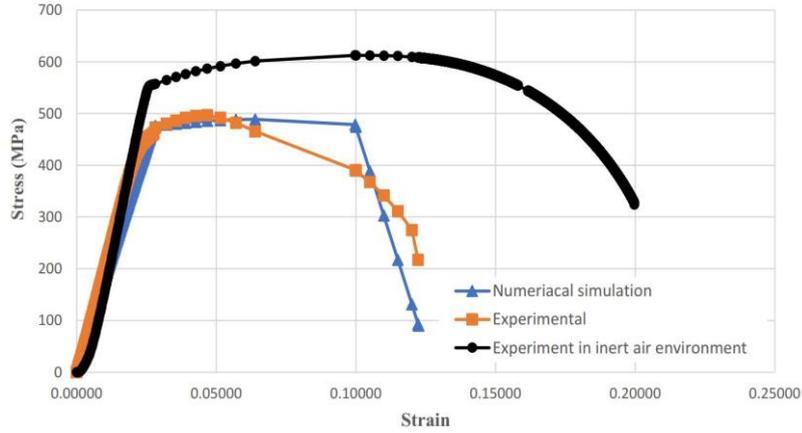

**Fig. 5.** Stress-Strain Curves of tensile test for the experiment, simulation, and experiment in the inert air environment.

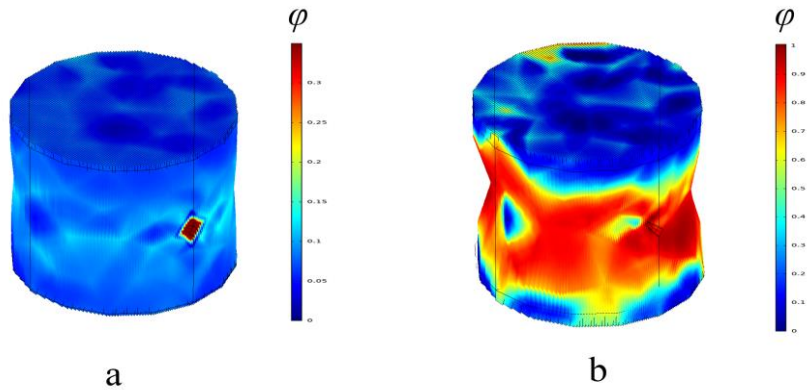

**Fig. 6**. The phase field damage growth due to SCC in the numerical model: (a) the initial distribution of damage on the surface of the rod, (b) the final damage distribution of the rod.

The phase field contour of the numerical model is shown in Fig. 6. The crack initiation begins uniformly from the pit, followed by crack propagation in the mid-section, ultimately resulting in the sample failure in agreement with the experiment [1].

### 4.3. SCC in a Notched Square Plate

This example is presented to study the effects of the interface kinetic coefficient, $L_{SCC}$ and the prescribed displacement (i.e., mechanical load) on the SCC failure. In the first part, by varying the $L_{SCC}$ parameter, its effects on crack initiation and growth are evaluated under constant boundary displacements. In the second part, the $L_{SCC}$ parameter is kept constant, and the effects of varying mechanical loads on the fracture behavior is investigated.

The problem consists of a $1\ mm \times 1\ mm$ square plate in the plane strain state with an edge notch of 0.5 mm in length and width of 0.01 mm, located in the mid-height of the plate. The bottom face of the plate is fixed clamped, and the top face is fixed in x-direction and subjected to prescribed displacement of in y-direction. In this model, $\sigma_y = 535\ MPa$ and $L = 0.0000235\ \frac{mm^2}{N.s}$, so based on equation (27), $L_{cm}$ is $0.0125\ \frac{1}{s}$, and other material properties are the same as the previous example. The structured mesh of the plate consists of 38,000 bilinear four-node fully-integrated elements.

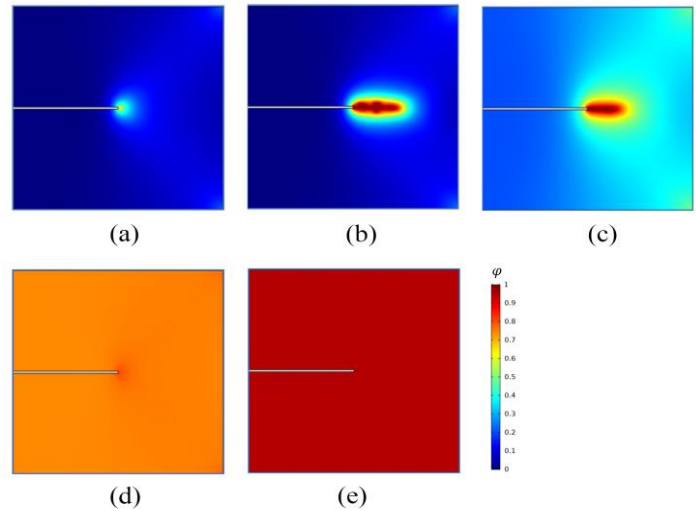

**Fig. 7.** Phase-field damage contours at the end of 5 days for different values of $L_{SCC}$: (a) $0.001\ \frac{mm^2}{N\cdot s}$, (b) $0.1\ \frac{mm^2}{N\cdot s}$, (c) $1\ \frac{mm^2}{N\cdot s}$, (d) $10\ \frac{mm^2}{N\cdot s}$, and (e) $100\ \frac{mm^2}{N\cdot s}$.

In the first part, the top face is fixed at 0.0055 mm displacement and the interface kinetic coefficient values vary as: $L_{SCC} = 0.001, 0.1, 1, 10$ and $100\ \frac{mm^2}{N.s}$. Fig. 7 shows the corresponding phase filed damage contours after 5 days. As observed in Fig. 7a, when $L_{SCC} =$ 0.001, the corrosion process is protracted and the synergy between the mechanical and electrochemical driving forces is negligible; thus, since mechanical load is low, no significant SCC is also observed. As Fig. 7b shows, by increasing $L_{SCC}$, a more corrosive environment is modeled and the interaction between the mechanical and electrochemical contributions is enhanced; thus, a crack nucleates from the notch tip and grows into the mid-width of the plate. By further increase of $L_{SCC}$ to 1 (Fig. 7c), along with the SCC crack growth, damage diffusion due to the corrosion is observed in the crack vicinity and the right corners of the sample where stress concentration exists. By increasing the $L_{SCC}$ to 10 and 100 in Figs. 7d and 7e again, the corrosion dominates the SCC. This is due to the acceleration of the corrosion process and the excessive contribution of the electrochemical part compared to the mechanical part in the formulation; in fact, the damage by corrosion diffuses into the entire domain and SCC does not occur. This shows that an appropriate value of the $L_{SCC}$ parameter is necessary to model the interaction between the electrochemical and mechanical parts and thus the SCC phenomenon. Section 3.3, explains the calibration process of this value based on the strain-stress curve of a sample in the desired corrosive environment.

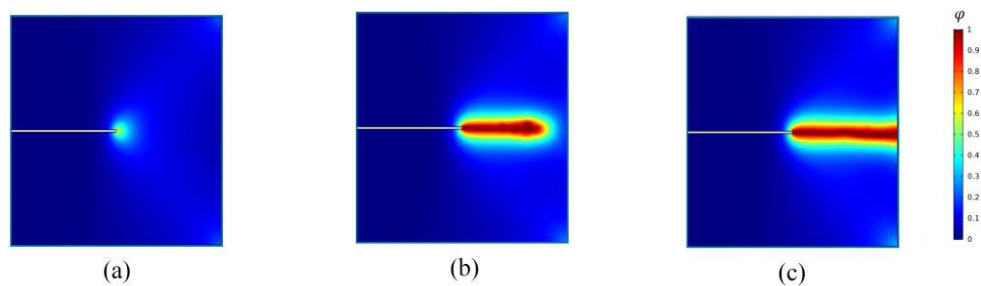

**Fig.8.** Phase-field damage contours a in the square sample at the end of 5 hrs at prescribed displacements: (a) 0.0055 mm, (b) 0.006 mm, and (c) 0.0065 mm.

In the second part, the interface kinetic coefficient is kept constant as the minimum value of the previous part, $L_{SCC} = 0.001 \frac{mm^2}{N \cdot s}$, and the prescribed displacements is varied as 0.0055, 0.006 and 0.0065 mm; after the displacement reaches the final value in $10\ s$, it remains unchanged in the course of analysis. Fig. 8 presents the phase-field damage contour at the end of the simulation ($t = 5$ hrs.). It is observed that for the applied displacement of 0.0055 mm, due to the low mechanical effects, the interaction between the mechanical and electrochemical parts is negligible, and SCC does not occur. With the increase of prescribed displacement, the SCC crack growth occurs from the notch tip and the greater the displacement, the more significant the crack growth (Fig. 8b and c).

**4.4. SCC failure in an end-plate connection**

In the next example, the application of the proposed formulation in a practical engineering problem is illustrated. For this purpose, the SCC failure of a pretensioned bolt in an end-plate connection of a cantilevered plate-girder beam to a box column is investigated. The connection assembly is shown in Fig. 9, and the dimensions are reported in table 2. Six M20 high strength bolts of iso 8.8 type are used in the slip critical connection with pre-tensioning forces of 150 kN. To reduce the computational costs, only one flange of the box column is modeled; the four edges of the column flange are restrained and a friction coefficient of $\mu = 0.15$ is assumed between the column flange and the end-plate. The diameter of the plate holes is 22 mm and the contact between bolt shafts and holes is neglected. The beam length is 1500 mm and it is subjected to a downward vertical displacement rate of $4.5 \times 10^{-4}\ \frac{mm}{s}$ at its cantilever end to model the mechanical contribution to SCC. All components are made of steel with material properties: $\sigma_y = 635\ MPa$, $E = 200\ GPa$, $v = 0.3$, and $\rho = 7850\ \frac{kg}{m^3}$. The phase-field SCC formulation is only applied to bolt #3 (Fig. 9a), with $L = 0.001\ \frac{mm^2}{N \cdot s}$, $L_{cm} = 0.635\ \frac{1}{s}$ and $L_{SCC} = 10\ \frac{mm^2}{N \cdot s}$. The fracture toughness and the regularization parameters are similar to example 4.2. At the beginning of the modeling, a pre-tensioning step is performed for the bolts; the mechanical and electrochemical loads are applied next.

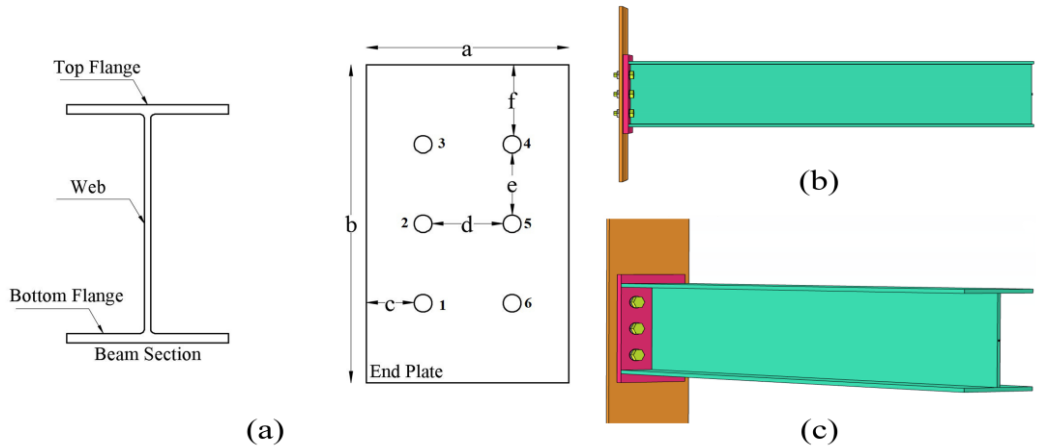

**Fig. 9.** Geometry of the end-plate connection assembly: (a) beam section and end-plate, (b) side view of the assembly, (c) 3D view of the assembly.

**Table 2:** The dimensions of the end-plate connection assembly.

| Parameter | value | Unit |
| --- | --- | --- |
| a | 250 | mm |
| b | 400 | mm |
| c | 59 | mm |
| d | 88 | mm |
| e | 78 | mm |
| f | 89 | mm |
| Hole diameter | 22 | mm |
| Beam top and bottom flange plates | 200×12 | mm |
| Beam web | 276×8 | mm |

Figure 10 shows the phase-field damage contours in bolt #3 at different time steps. It is observed that as the applied displacement increases, the damage grows from the tensile region of the bolt at the shaft and head junction, and spreads to the whole section. The damage growth accelerates by the simultaneous effects of corrosion and stress concentration due to the reduction of the bolt cross-sectional area. Eventually, the bolt fractures from the head junction in a brittle manner.

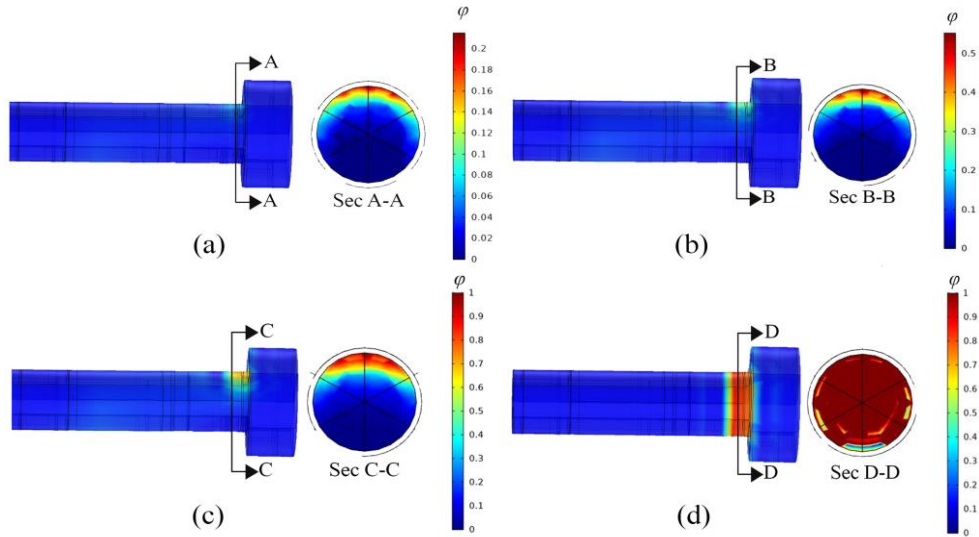

**Fig. 10**. SCC crack growth in bolt #3 at different times: (a) 8640 second, (b) 11232 s, (c) 12096 s, and (d) 12960 s.

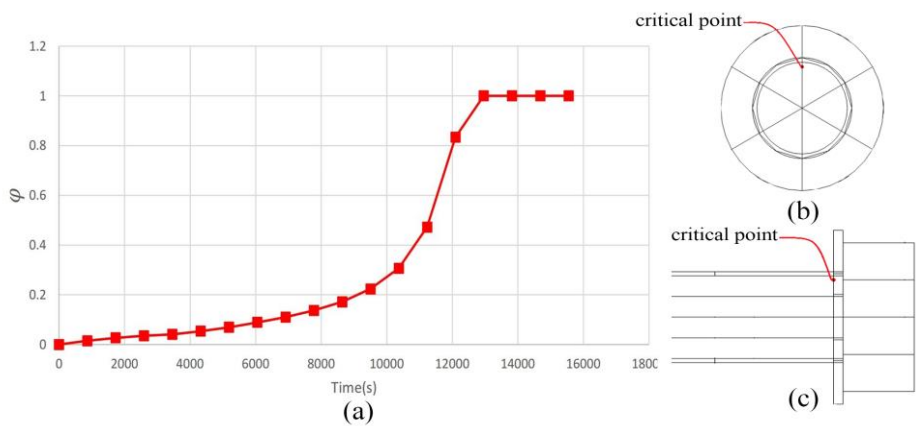

**Fig. 11.** The phase-field parameter in a critical point: (a) the Phase field parameter time-history, (b) the location of the critical point in the front view, (c) the critical point in the side view.

Fig. 11 shows the phase field damage time history at a critical point in the tensile side of the bolt. The brittle and sudden nature of the SCC failure is evident in this diagram. In the initial stages, the slope of the graph, i.e., damage growth rate, is small, but at the time of failure, this slope rises suddenly.

Figure 12 displays the von Mises stress contour resulting from the bolt pre-tensioning and the contact between the bolt heads and the end-plate. Initially, the pre-tensioning force generates a stress of 320 MPa at the contact area. However, as the analysis proceeds, the top bolts extend and the end-plate separates from the column flange due to the applied displacement at the beam end. Consequently, there is stress redistribution within the joint. Bolt #3, which is susceptible to SCC, experiences failure and separates from the end-plate, causing the stress to approach zero. In contrast, bolt #4 undergoes a higher level of stress and bears more load due to stress redistribution. The stress contour on the deformed configuration of the end plate is depicted in Figure 13, illustrating one load step before failure and at the time of bolt failure. The contour effectively captures the separation of the upper end plate section from the column flange due to tension. In the final stage, the consideration of SCC in bolt #3 leads to its failure and subsequent separation from the end-plate. As a result, the stress in the affected bolt and its surrounding region in the end-plate reduces to zero.

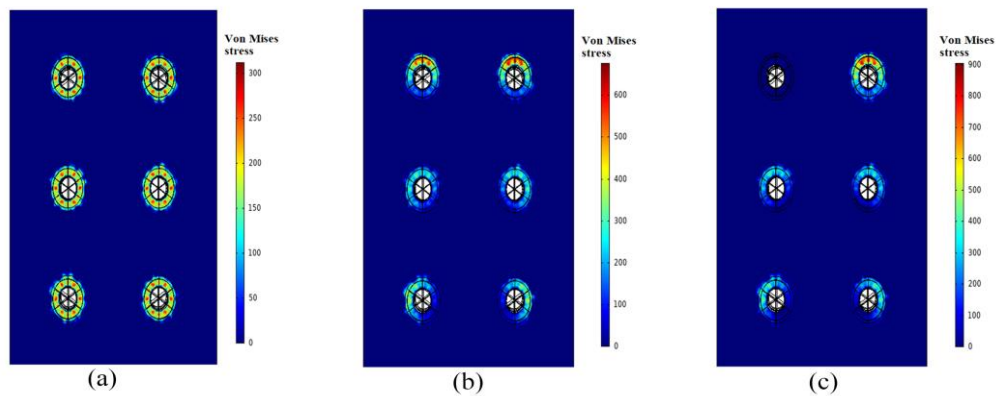

**Fig. 12.** The von Mises stress contour due to the pre-tensioning force and the contact between the bolt head and the end-plate at different time steps: (a) at the beginning of the analysis, (b) 8640 s, (c) 12960 s.

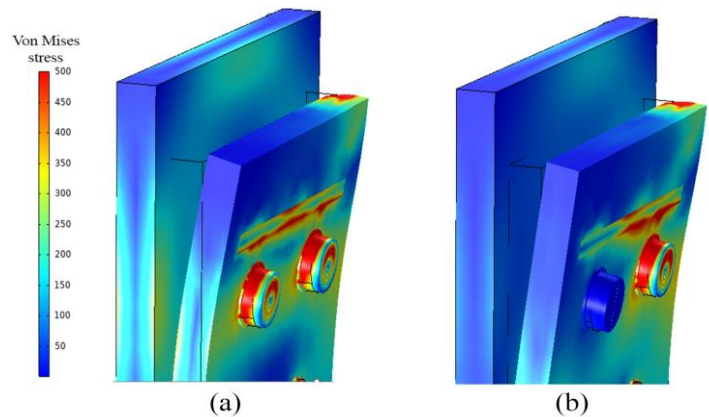

**Fig. 13.** The von Mises stress contour on the deformed configuration of the end-plate (a) one load step before the failure in 12096 s , (b) at the failure in 12960 s.

Figure 14 presents a comparison of the moment-rotation diagrams for the connection assembly with and without SCC effects. The diagram shows that the connection's capacity decreases with the failure of the bolt at t=12960 s, and a sudden drop in the moment-rotation curve is observed. Consequently, the connection cannot reach its ultimate capacity as per the initial design, and the deterioration of one of the bolts due to SCC leads to a reduction in the connection's capacity.

The figure demonstrates that, at the end of the analysis, the connection can withstand a lower load compared to the scenario where the beam is not exposed to the corrosive environment. In summary, the results suggest that SCC can significantly impact the structural performance of bolted connections, leading to a decrease in their load-carrying capacity and potentially compromising their safety and reliability. Therefore, adequate measures must be taken to mitigate the effects of SCC and prevent its occurrence in bolted joints.

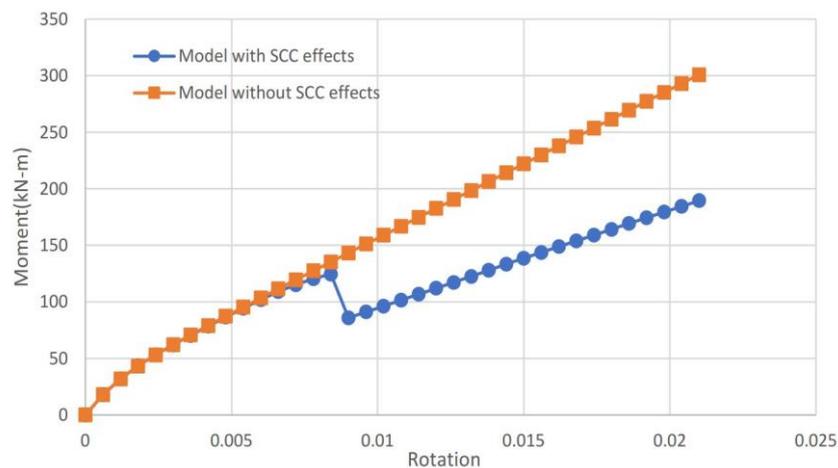

**Fig. 14.** Comparison between the moment-rotation diagrams of the model without SCC (orange) and the model with SCC (blue).

## 4.5. SCC in a gusset plate connection

The last example is devoted to the study of the failure development in a gusset plate connection of a practical problem. Three cases of pure corrosion, pure mechanical loading, and SCC are considered. Lastly, a parametric study is performed on $L_{cm}$ to assess its effects on SCC failure simulation. The problem consists of a rigid beam connected to a column with a gusset plate similar to the geometry presented in reference [34]. Figure 15 shows the connection assembly, and the model dimensions are presented in Table 3. To reduce the computational costs, only the flange of the column is modelled; the four edges of the column flange are restrained. Eight M25 high-strength bolts of iso 8.8 type are used as the connectors and the diameter of the holes is assumed 1.8 mm larger than the diameter of the bolts. To have a slip-critical connection, a pre-tensioning force of 150 kN has been applied to the bolts. A constant displacement rate of $4 \times 10^{-4} \frac{mm}{s}$ is exerted to the upper edges of the rigid beam and its end face in the longitudinal direction. The contact between the lower surface of the bolt head and the upper surface of the washer is considered to be of a "continuous" type and the contact surface between the lower surface of the washer and the upper surface of the gusset plate is considered to be of "friction" type with a friction coefficient of $\mu = 0.2$. Also, a contact surface is defined between the bolt shaft and the inner surface of the hole to prevent the bolt penetration into the gusset plate in the case of bolt slip. All components are made of steel with material properties similar to the section 4.4, except that $\sigma_y = 325\ MPa$, $L_{SCC} = 1\ \frac{mm^2}{N \cdot s}$, $L = 0.001\ \frac{mm^2}{N \cdot s}$, and $L_{cm} = 0.325\ \frac{1}{s}$. The phase-field damage is only applied to the gusset plate, and other components remain intact during the analysis. Quadratic brick elements are used in the FEM model, and different mesh sizes are applied to each part of the model to increase the accuracy and reduce the computational costs. The analysis is performed in two stages: first, the bolts are pre-tensioned, and then the model is analyzed under the applied mechanical and\or corrosion loads. Full Newton-Raphson method is used to solve the nonlinear system of equations.

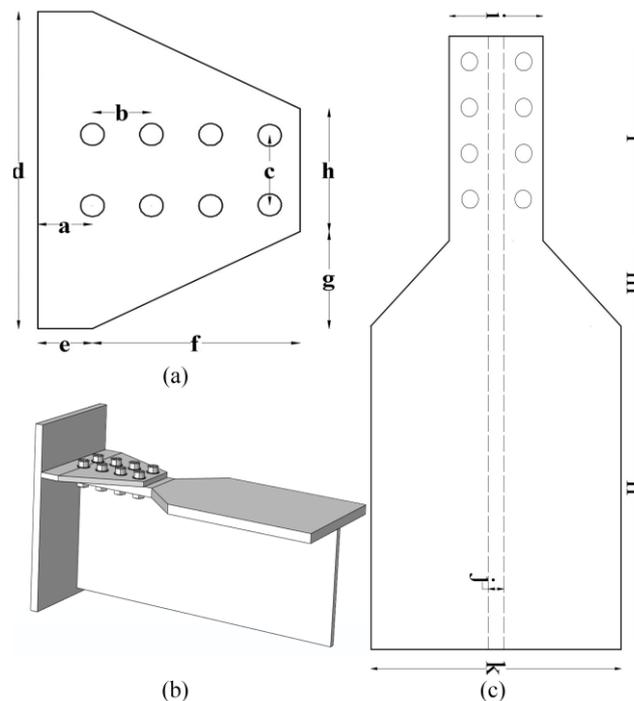

**Fig.15**. The gusset plate connection assembly: (a) the gusset plate, (b) the 3D view of the assembly, and (c) the rigid beam

Table 3. The Dimensions of the gusset plate connection.

| Parameter | value (mm) |
|---|---|
| a | 63.8 |
| b | 68.3 |
| c | 86.2 |
| d | 393.7 |
| e | 63 |
| f | 239.7 |
| g | 120.7 |
| h | 152.4 |
| Thickness of the gusset plate | 14.3 |
| i | 152.4 |
| j | 25.4 |
| k | 406.5 |
| l | 305 |
| m | 127.3 |
| n | 482 |
| Thickness of the rigid beam | 25.4 |
| Height of the rigid beam | 305 |

In the first part, the pure corrosion process is investigated, neglecting any mechanical effects, i.e. $L_{cm} = 0$. The electrochemical formulation is only applied to a rectangular zone in the middle of the gusset plate to simplify the model and reduce the computational cost. The corrosion Dirichlet boundary conditions, i.e. $c = 0$ and $\phi = 1$, are applied to the inside surfaces of the holes of the gusset plate. As shown in Fig. 16, the corrosion failure starts from the holes and diffuses uniformly to the gusset plate. As expected, due to the absence of the mechanical effects, the pit-to-crack transformation and crack branching phenomena do not occur.

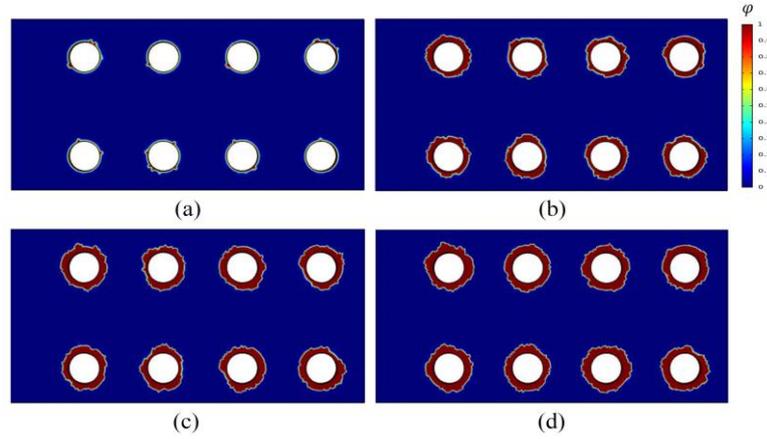

**Fig. 16**. Phase-field damage growth due to pure corrosion at: (a) 1 day, (b) 10 days, (c) 15 days, and (d) 20 days.

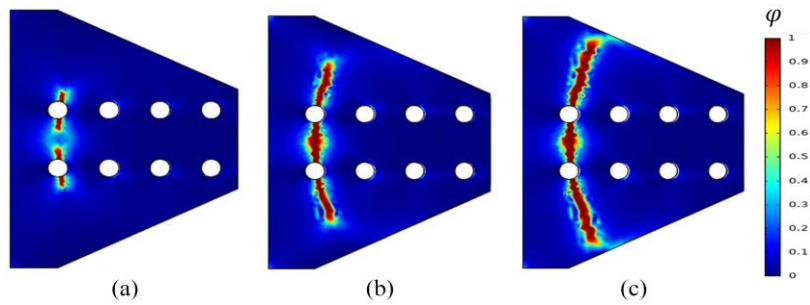

**Fig. 17.** Phase-field damage growth due to pure mechanical failure at: (a) 0.25 days, (b) 0.31 days and (c) 0.36 days.

In the second part, the pure mechanical failure of the gusset plate, i.e. $L_{SCC} = 0$, is investigated. To this end, the model is first subjected to the bolts pre-tensioning forces, and then the prescribed displacement rate is applied. As Figure 17 shows, when the displacement increases, cracks gradually nucleate from the two left end holes at the stress concentration zones; these cracks propagate in a direction perpendicular to the applied displacement, coalesce and eventually advance toward the top and bottom edges of the gusset plate. This is in agreement with the Whitmore section theory [35] for gusset plate design which states that the forces in a gusset plate spread on an effective cross sectional area found by the intersection of two lines drawn at an angle of 30 degrees from the first connector row with the line drawn at the last connector row; the failure due to yielding or fracture limit states in the gusset plate may occur at this critical cross-section, as is the case in this simulation (Fig. 17).

In the third part, the phenomena of stress corrosion cracking (SCC) is considered. In this case, the interaction between mechanical and electrochemical effects causes crack nucleation and growth in the sample, thus, the crack growth pattern is different from the previous two parts. Figure 18 shows the SCC phase-field damage evolution in the course of the analysis. Although, similar to the pure mechanical case, cracks initially form around the end holes due to the high-stress concentration, with the continuation of the analysis, further cracks appear in other connection holes. Also, the cracks grow obliquely with respect to the loading direction and eventually coalesce to cause the complete failure of the gusset plate. This clearly shows the importance of the consideration of the interaction between mechanical and electrochemical effects in the simulation of SCC phenomena.

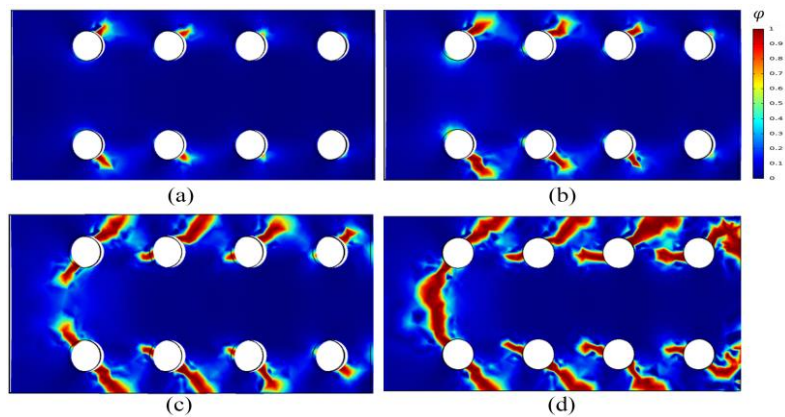

**Fig. 18.** Phase-field damage growth due to SCC for $L_{SCC} = 1\frac{mm^2}{N \cdot s}$ and $L_{cm} = 0.325\frac{1}{s}$ at: (a) 0.40 days, (b) 0.45 days, (c) 0.48 days, and (d) 0.49 days.

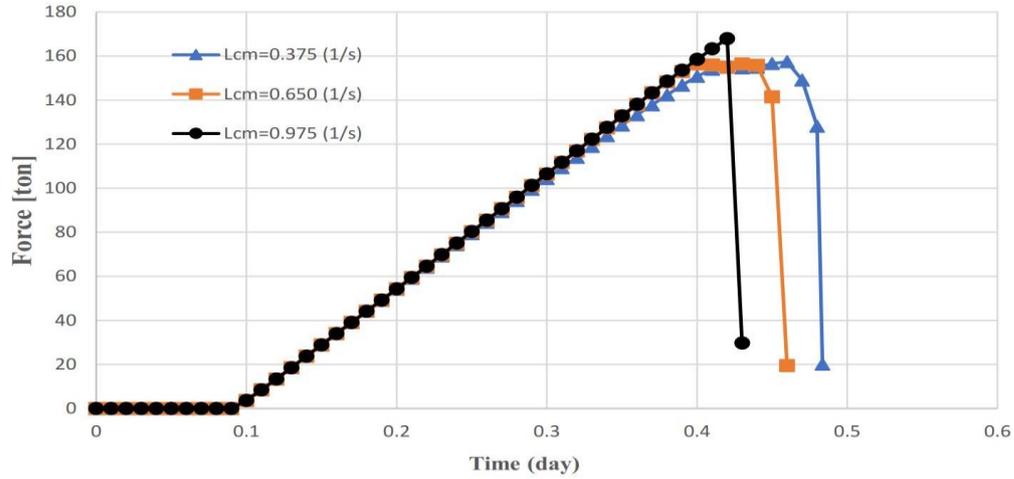

**Fig. 19.** Effect of different values of $L_{cm}$ on time to failure of the gusset plate due to SCC

In the last section, a parametric analysis is performed on $L_{cm}$ to investigate how changes in the concentration of the corrosive environment affect the failure of the gusset plate. By increasing the corrosion current of the electrolyte, the rate of the reaction at the corrosion boundary is increased, which in turn causes an increase in the value of $L$ and accelerates the sample failure. To simulate this effect, $L$ values of 0.001, 0.002, and 0.003 $\frac{mm^2}{N \cdot s}$ are considered, which correspond to $L_{cm}$ values of 0.325, 0.650, and 0.975 $1/s$, respectively, according to Equation (27). The force time-histories for different values of $L_{cm}$ are compared in Figure 19. It is observed that the time to failure of the gusset plate is decreased and the sample fails in a more brittle manner with an increase in $L_{cm}$. The damage contours at the time of failure and one time step before the failure are presented in Figures 20. It is evident that damage grows more abruptly with an increase in $L_{cm}$.

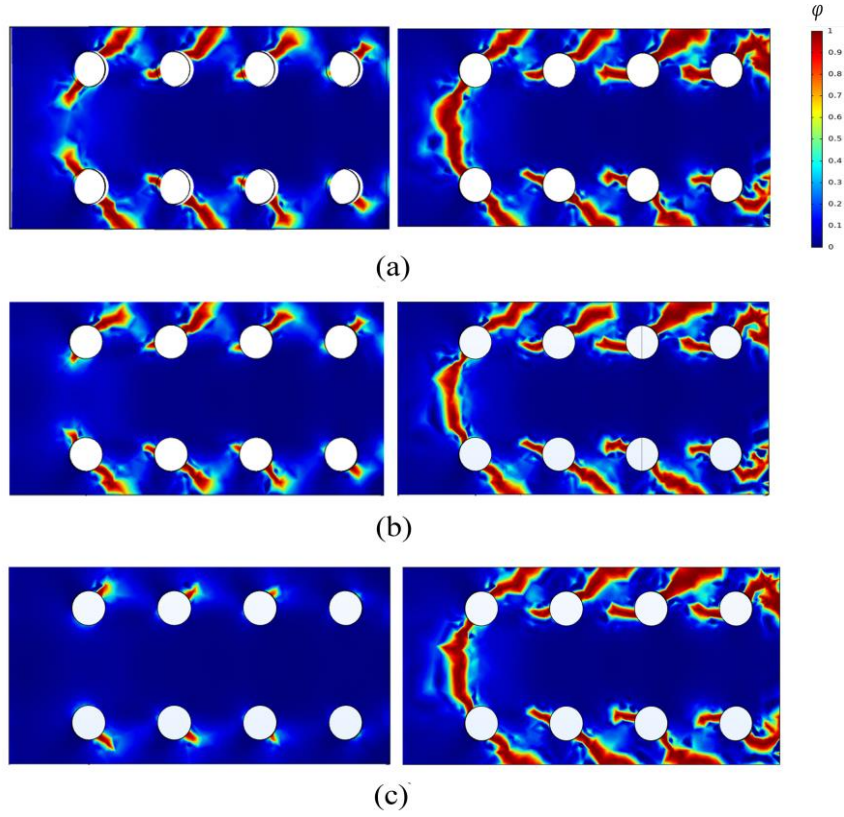

**Fig. 20.** Phase-field damage growth due to for $L_{SCC} = 1 \frac{mm^2}{N \cdot s}$ and prescribed displacement $4 \times 10^{-4} \frac{mm}{s}$ at (a) $L_{cm} = 0.325 \frac{1}{s}$ (Left 0.48 day, right 0.49 day), (b) $L_{cm} = 0.650 \frac{1}{s}$ (Left 0.44 day, right 0.45 day) (c) $L_{cm} = 0.975 \frac{1}{s}$ (Left 0.42 day, right 0.43 day)

## 5. Conclusions

This study introduces a novel framework for simulating Stress Corrosion Cracking (SCC) using the phase field method. The proposed phase-field parameter effectively combines the effects of both mechanical and electrochemical damages and is determined by considering the contributions of mechanical and electrochemical energies and their variations. Unique interfacial kinetic coefficients are suggested for each energy contribution, allowing for simulation of the impact of mechanical and electrochemical factors and providing greater flexibility in modeling different materials and corrosive environments. The governing equations of the proposed formulation include the modified Allen-Cahn equation for the phase-field parameter, the Cahn-Hilliard equation for corrosion ion concentration, and the equilibrium equations for the solid body. This formulation can simulate a range of scenarios, including pure corrosion, pure mechanical damage, and combined mechanical and corrosion damages. The proposed formulation was implemented in COMSOL-Multiphysics software, and several numerical examples were provided to demonstrate the robustness and accuracy of the method. The results indicate that the proposed method can be effectively used to model SCC failure in engineering steel structures.